\documentclass[runningheads]{llncs}

\usepackage[colorlinks=true, urlcolor=blue, linkcolor=blue, citecolor=blue]{hyperref}
\usepackage{orcidlink}
\usepackage{float}
\usepackage{marvosym}

\usepackage{enumitem}

\usepackage[T1]{fontenc}
\usepackage{graphicx}
\usepackage{amsmath, amssymb} 
\usepackage{booktabs}  

\usepackage{subfig}
\usepackage{caption}
\usepackage{fontawesome}

\begin{document}
\title{AutoFSM: A Multi-Agent Framework \\for FSM Code Generation\\with IR and SystemC-Based Testing}
\titlerunning{AutoFSM}
%
\author{Qiuming Luo\inst{1}\orcidlink{0000-0003-3622-5386} \and
Yanming Lei\inst{1}\textsuperscript{(\Letter)} \and
Kunzhong Wu\inst{1} \and \\
Yixuan Cao\inst{1}\orcidlink{0009-0006-6241-4251} \and
Chengjian Liu\inst{2}
}
\authorrunning{Q. Luo et al.}
%
\institute{The College of Computer Science and Software Engineering, Shenzhen University, Shenzhen 518000, China\\
\email{lqm@szu.edu.cn, leiyanming2023@email.szu.edu.cn, wukunzhong2019@email.szu.edu.cn, caoyixuan2019@email.szu.edu.cn}\\
 \and
College of Big Data and Internet, Shenzhen Technology University, \\Shenzhen 518000, China\\
\email{liuchengjian@sztu.edu.cn}
}
\maketitle              
\begin{abstract}
With the rapid advancement of large language models (LLMs) in code generation, their applications in hardware design are receiving growing attention. However, existing LLMs face several challenges when generating Verilog code for finite state machine (FSM) control logic, including frequent syntax errors, low debugging efficiency, and heavy reliance on test benchmarks. To address these challenges, this paper proposes AutoFSM, a multi-agent collaborative framework designed for FSM code generation tasks. AutoFSM introduces a structurally clear intermediate representation (IR) to reduce syntax error rate during code generation and provides a supporting toolchain to enable automatic translation from IR to Verilog. Furthermore, AutoFSM is the first to integrate SystemC-based modeling with automatic testbench generation, thereby improving debugging efficiency and feedback quality. To systematically evaluate the framework's performance, we construct SKT-FSM, the first hierarchical FSM benchmark in the field, comprising 67 FSM samples across different complexity levels. Experimental results show that, under the same base LLM, AutoFSM consistently outperforms the open-source framework MAGE on the SKT-FSM benchmark, achieving up to an 11.94\% improvement in pass rate and up to a 17.62\% reduction in syntax error rate. These results demonstrate the potential of combining LLMs with structured IR and automated testing to improve the reliability and scalability of register-transfer level (RTL) code generation.
\keywords{Large language model  \and Multi-agent framework \and Finite state machine \and Code generation.}
\end{abstract}
\section{Introduction}
In modern digital system design, the finite state machine (FSM) \cite{minns2008fsm} continues to play an irreplaceable role as a core control architecture. From communication protocol parsing to processor instruction scheduling, from embedded control to interface timing management, FSM provides structured and standardized control logic frameworks for digital systems through their state transition mechanisms. However, as the number of states increases, the design complexity and the workload of writing hardware description language (HDL) code (such as Verilog) also rise significantly. This not only increases the likelihood of human coding errors but also greatly extends development time. Although traditional Electronic Design Automation (EDA) tools have made progress in automatically generating Verilog code for FSMs, they still heavily rely on manual intervention.

In recent years, large language models (LLMs), e.g., DeepSeek \cite{liu2024deepseek} and GPT-4 \cite{achiam2023gpt}, have demonstrated powerful capabilities in understanding and generating natural language, opening new pathways for automatic code generation from natural language to hardware description. To fully exploit the potential of LLMs in hardware design, researchers have conducted various explorations. For example, Chip-Chat \cite{blocklove2023chip} designs an 8-bit microprocessor through dialogue with GPT-4, but the manual interaction process leads to low efficiency. VeriGen \cite{thakur2024verigen} fine-tune a CodeGen model using corpora collected from GitHub and textbooks, achieving better Verilog code generation performance than GPT-3.5-Turbo. However, its single-pass generation method struggles to ensure complete code correctness. VerilogCoder \cite{ho2025verilogcoder} further introduces a multi-agent mechanism capable of accurately locating faulty signals based on simulation errors, thereby enhancing the framework’s debugging capabilities. Nonetheless, its reliance on manually designed test programs in the benchmark limits its practical applicability.

To address these challenges, we propose the following solution. First, we construct a benchmark specifically for evaluating FSM code generation: SKT-FSM, which contains 67 FSM samples categorized into three levels of complexity—easy, medium, and hard. Second, we introduce AutoFSM, the first multi-agent framework dedicated to FSM-oriented hardware code generation. This framework incorporates a simplified intermediate representation (IR) and a supporting toolchain for automatically translating the IR into Verilog code, significantly reducing syntax error rates. Furthermore, AutoFSM is the first to integrate SystemC \cite{panda2001systemc} modeling and automated testbench generation. By introducing a differential testing strategy in the verification process, it greatly improves code validation and debugging efficiency.

Our contributions are summarized as follows:
\begin{itemize}
\item We construct SKT-FSM, the first hierarchical benchmark for FSM code generation, enabling systematic evaluation of FSM-oriented frameworks.
\item We propose a simple intermediate representation and developed a corresponding toolchain to convert it into Verilog code, effectively reducing the syntax error rate.
\item We are the first to integrate SystemC modeling and automatic test program generation into a hardware code generation framework, addressing the issue of relying on manually designed benchmark for error feedback and significantly improving the framework’s error correction capability.
\item On the SKT-FSM benchmark, AutoFSM consistently outperforms the open-source framework MAGE \cite{zhao2024mage} under the same base LLM, with up to 11.94\% higher pass rate and 17.62\% lower syntax error rate.
\end{itemize}

\section{Background and Motivation}
\subsection{Method of Generating RTL Code Based on LLM}
Research on applying LLMs to register-transfer level (RTL) design primarily focuses on three areas: (1) prompt engineering techniques, (2) model fine-tuning, and (3) autonomous agent-based methods that iteratively improve code generation by incorporating user-defined tool feedback \cite{pan2025survey}.

In terms of prompt engineering, ChipGPT \cite{chang2023chipgpt} leverages in-context learning to effectively improve the generation of Verilog code from natural language specifications. RTLLM \cite{lu2024rtllm} introduces a self-planning prompt engineering technique, significantly enhancing the capability of GPT-3.5 to generate RTL designs from natural language instructions.

In the domain of model fine-tuning, BetterV \cite{zehua2024betterv} applies instruction fine-tuning to LLMs on domain-specific datasets and introduces a generative discriminator to evaluate and filter candidate Verilog code. On the VerilogEval benchmark, this method outperforms GPT-4 in both accuracy and code quality.

While prompt optimization and model fine-tuning have positively contributed to enhancing LLMs' RTL code generation capabilities, they still rely heavily on manual effort during simulation and verification, limiting the realization of a fully integrated design-verification workflow. To overcome this, researchers have introduced memory management, auxiliary toolchains, and automatic feedback mechanisms to construct autonomous agent systems with the ability for self-planning and self-improvement. However, in single-agent settings where all tasks share the same context, the agent's ability to generate accurate code and correct errors tends to degrade over time.
This drives current research toward multi-agent collaboration frameworks to enhance system robustness and scalability.

MAGE \cite{zhao2024mage} is the first open-source multi-agent RTL generation framework. It employs a high-temperature candidate sampling strategy to improve code diversity and integrates a checkpoint-based error localization mechanism to enhance Verilog code repair accuracy. On the VerilogEval-Human v2 benchmark \cite{pinckney2024revisiting}, MAGE achieves a pass rate of 95.7\%.

\subsection{Motivation}
Although the aforementioned methods have made significant progress in improving LLMs' ability to generate RTL code, several critical challenges remain. Most existing frameworks adopt a direct Verilog generation approach. However, research \cite{tsai2024rtlfixer} has shown that 55\% of the errors produced by LLMs when generating Verilog are syntax-related. In systems equipped with error feedback mechanisms, such syntax errors often intermingle with functional errors, severely hindering the model’s ability to localize and correct functional issues. The root cause lies in the lack of sufficient Verilog training data in current LLMs, especially when compared to mainstream programming languages such as Python or C. As a result, LLMs have a limited understanding of Verilog’s syntactic structures and semantic constraints. To address this issue, we propose replacing direct Verilog generation with a simpler intermediate representation that is both semantically explicit and well-supported by training data. This representation is structurally clear and semantically unambiguous, and can be reliably compiled into valid Verilog using a custom toolchain, thereby significantly reducing syntax error rates and enhancing the engineering usability of the framework. Design details are provided in Section~\ref{sec:IR}.

Moreover, previous studies \cite{tsai2024rtlfixer} have shown that LLMs often fail to produce syntactically and functionally correct hardware code in a single pass. Therefore, RTL code generated by LLMs must be validated through testbench simulation and iteratively refined based on simulation results—an approach consistent with how human engineers design hardware. However, multi-agent frameworks such as \cite{ho2025verilogcoder,zhao2024mage} typically rely on predefined test programs from a benchmark for error feedback. These programs often provide limited diagnostic information, making it difficult for LLMs to accurately identify the root causes of errors, and thus limiting the effectiveness of debugging and repair. Furthermore, these frameworks lack the ability to generate test programs automatically, relying instead on human-written golden designs and verification logic, which restricts their applicability in real-world scenarios. To overcome this limitation, we propose integrating an automated test program generation mechanism that can dynamically generate validation logic for each design description. This not only improves the interpretability of error localization, but also enhances the generalizability and scalability of the framework across various task scenarios. Design details are discussed in Section~\ref{sec:test}.

Finally, existing RTL code evaluation benchmarks also exhibit notable limitations. The most widely used benchmark, VerilogEval [8], includes only 26 FSM-related tasks, accounting for just 15.4\% of the benchmark. Moreover, the FSM examples in this benchmark are all under 60 lines of code, with simple structures that fail to fully reflect the complexity of real-world FSM control logic. To address this shortcoming, we further develop a dedicated benchmark tailored for FSM generation tasks, enabling more accurate evaluation of different frameworks’ capabilities in modeling complex control logic. Details of the benchmark construction are presented in Section~\ref{sec:benchmark}.

\section{Benchmark Construction}\label{sec:benchmark}
\begin{figure}[htbp]
	\centering
	\includegraphics[width=0.9\textwidth]{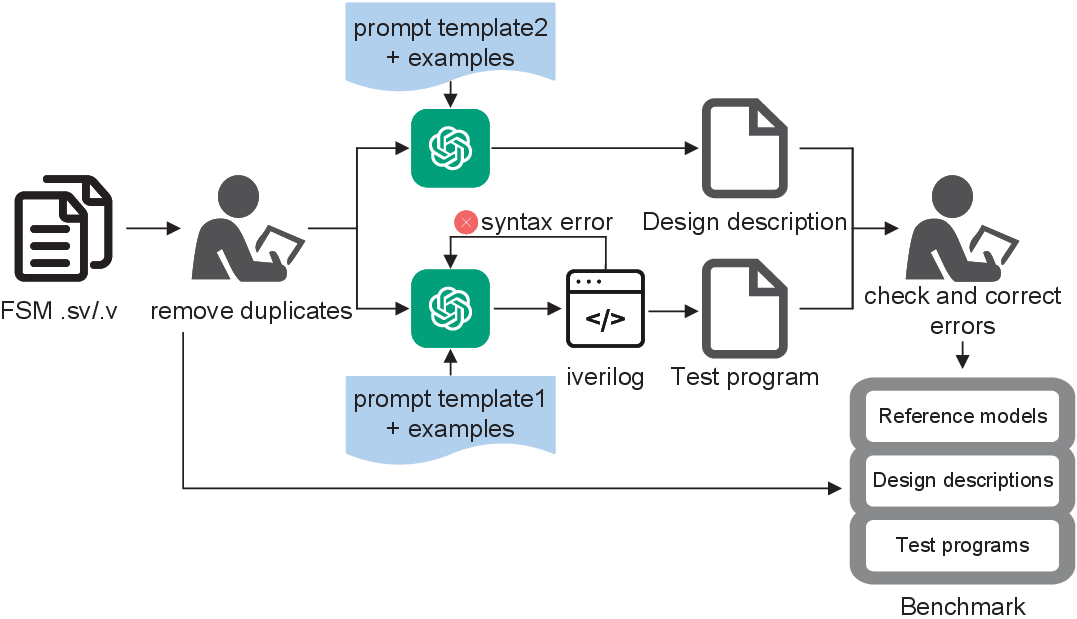}
	\caption{Workflow for benchmark construction}
	\label{benchmark}
\end{figure}

In this work, we first construct a benchmark specifically designed to evaluate the FSM code generation capabilities of various frameworks. The benchmark consists of three main components: design descriptions, reference models, and test programs, and follows the organizational structure of the VerilogEval v2 benchmark to ensure compatibility and ease of adoption by other frameworks. We use the 100 open-source FSMs released in \cite{bhandari2024llm} as our initial data source. To improve the efficiency of benchmark construction, we introduce LLM-assisted processing, as illustrated in Fig.~\ref{benchmark}. First, we manually review and remove redundant or structurally invalid FSM samples, ultimately retaining 67 representative FSMs as the reference models in the benchmark. Next, we input the RTL code of each reference model into GPT-4 and use a customized prompt template to automatically generate the corresponding design description. For test program generation, GPT-4 first produces an initial version based on the reference model. We then perform syntax validation using the Icarus Verilog (iverilog) compiler. If any syntax errors are detected, the error messages are fed back into GPT-4 for iterative correction, until a syntactically valid test program is obtained. All design descriptions and test programs generated by the LLM are further subjected to manual review and verification to ensure semantic alignment and functional correctness with respect to the reference models. If any issues are identified, manual corrections are performed.

After the benchmark is finalized, we perform statistical analysis on three key metrics for each sample: lines of code, number of states, and number of state transitions. To standardize the complexity across different samples, we apply normalization using equations~\eqref{equation:line}–\eqref{equation:edge}. A composite complexity score is then calculated using equation~\eqref{equation:score}.

\begin{equation}
L' = \frac{L_{i}}{L_{\text{max}}} \label{equation:line}
\end{equation}
\begin{equation}
S' = \frac{S_{i}}{S_{\text{max}}} \label{equation:state}
\end{equation}
\begin{equation}
E' = \frac{E_{i}}{E_{\text{max}}} \label{equation:edge}
\end{equation}
\begin{equation}
\text{score} = 0.3 \times S' + 0.5 \times E' + 0.2 \times L' \label{equation:score}
\end{equation}
where $L_i$, $S_i$, and $E_i$ represent the line count, number of states, and number of state transition edges of the $i$-th sample, respectively. $L_{\max}$, $S_{\max}$, and $E_{\max}$ denote the maximum values of line count, state count, and edge count across all samples. $L'$, $S'$, and $E'$ are the normalized values of the $i$-th sample for each respective metric. $\text{score}$ represents the overall complexity score of the $i$-th sample.
Based on the resulting scores, we categorize the samples into three difficulty levels: easy, medium, and hard, as shown in Table~\ref{tab:sample-count}.

\begin{table}[htbp]
\
  \centering
  \caption{Sample count by difficulty level}
    \begin{tabular}{ccc}
    \toprule
    \textbf{Difficulty level}  & \textbf{Count} & \textbf{Score}\\
    \midrule
    Easy    & 27 & $\text{score}<0.15$ \\
    Medium    & 27 & $0.15\le \text{score}<0.31$\\
    Hard     & 13 & $\text{score}\ge 0.31$\\
    \bottomrule
    \end{tabular}%
  \label{tab:sample-count}%
\end{table}%

Through this workflow, we have constructed a well-structured and comprehensive FSM code generation benchmark, which can serve as a standardized evaluation suite for assessing FSM generation capabilities in future research.

\section{AutoFSM Design}
\subsection{AutoFSM Overview}\label{sec:overiew}
\begin{figure}[H]
\centering
\subfloat[\centering Agents with different division of labor in AutoFSM]{\label{fig:agents}\centering\includegraphics[width=0.8\textwidth]{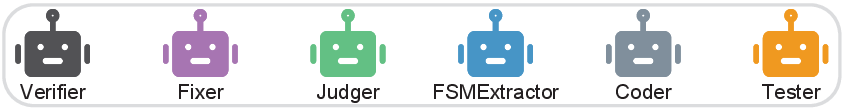}}\hfill
\subfloat[\centering Flow overview of AutoFSM]{\label{fig:flow}\centering\includegraphics[width=0.8\textwidth]{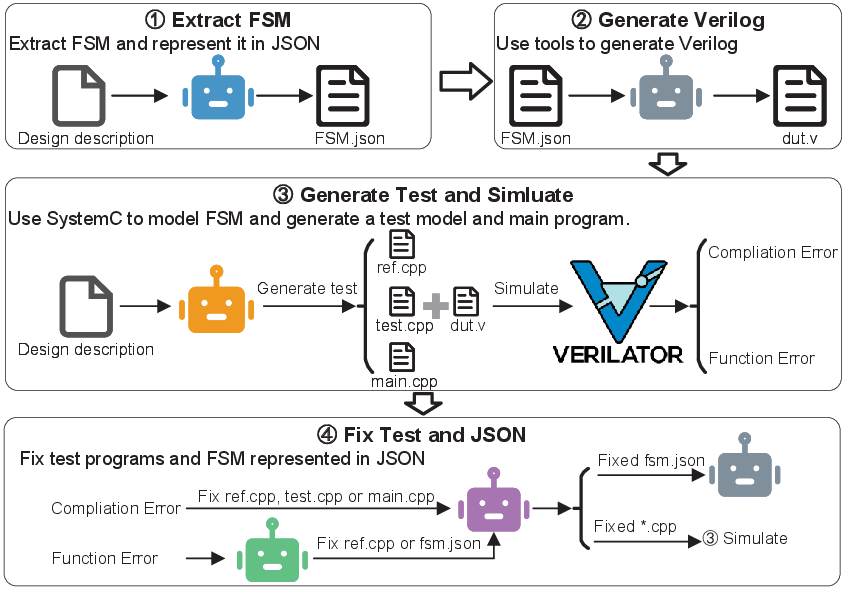}}\hfill
\subfloat[\centering JSON to Verilog code and simulation details]{\label{fig:detail}\centering\includegraphics[width=0.8\textwidth]{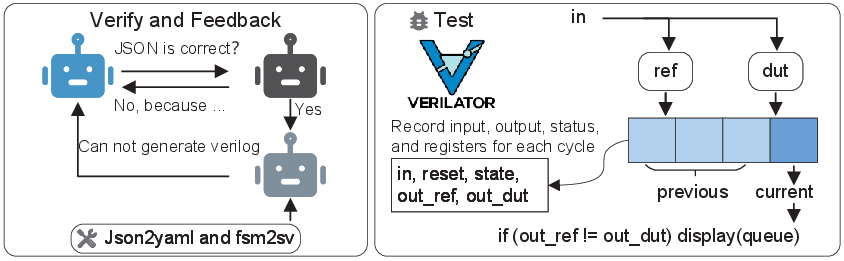}}\hfill
\caption{Overall architecture of AutoFSM}
\label{fig:overview}
\end{figure}

Fig.~\ref{fig:overview} illustrates the overall architecture of the AutoFSM framework. As shown in Fig.~\ref{fig:agents}, each agent within the framework is assigned a specific role: (1) The FSMExtractor agent parses the design description and converts it into a predefined JSON-format intermediate representation (IR). (2) The Verifier agent checks whether the IR accurately reflects the original design intent. (3) The Coder agent generates Verilog code based on the IR. (4) The Tester agent is responsible for generating test programs and conducting simulations. (5) The Fixer agent identifies and corrects errors in either the IR or the test programs. (6) The Judger agent analyzes simulation error logs to determine the root cause of the failure and provides guidance for targeted correction.

Based on the above agent responsibilities, we design an efficient collaborative workflow, as shown in Fig.~\ref{fig:flow}: Step 1: Extract the FSM from the design description and represent it in JSON format. Step 2: Generate Verilog code from the JSON representation. The details of these two steps are described in Section~\ref{sec:IR}. Step 3: Generate the testbench and simulate the Verilog code together with the test program to obtain simulation results. Step 4: If the simulation results indicate a compilation error, the Fixer agent is directly invoked to correct the issue. If a functional error is detected, the Judger agent first identifies the underlying cause, after which the Fixer agent performs targeted repairs accordingly. The details of the last two steps are stated in Section~\ref{sec:test}.

\subsection{Intermediate Representation for Error Reduction}\label{sec:IR}
Inspired by fsm2sv\footnote{\url{https://github.com/mohamed/fsm2sv}}—a tool capable of automatically generating Verilog code from FSM descriptions in YAML format—we propose a novel FSM design generation approach that integrates LLMs with fsm2sv. This approach introduces a structured IR, effectively decomposing the complex and error-prone task of Verilog code generation into two distinct stages. First, an LLM translates the natural language description of the FSM into a formal, machine-readable IR. Then, fsm2sv is used as a backend tool to automatically convert this representation into valid Verilog code. This method not only mitigates the high syntax error rate typically associated with direct Verilog generation by LLMs, but also leverages the LLM’s strengths in semantic comprehension and natural language-to-structured-data translation.

\begin{figure}[htbp]
	\centering
\includegraphics[width=\textwidth]{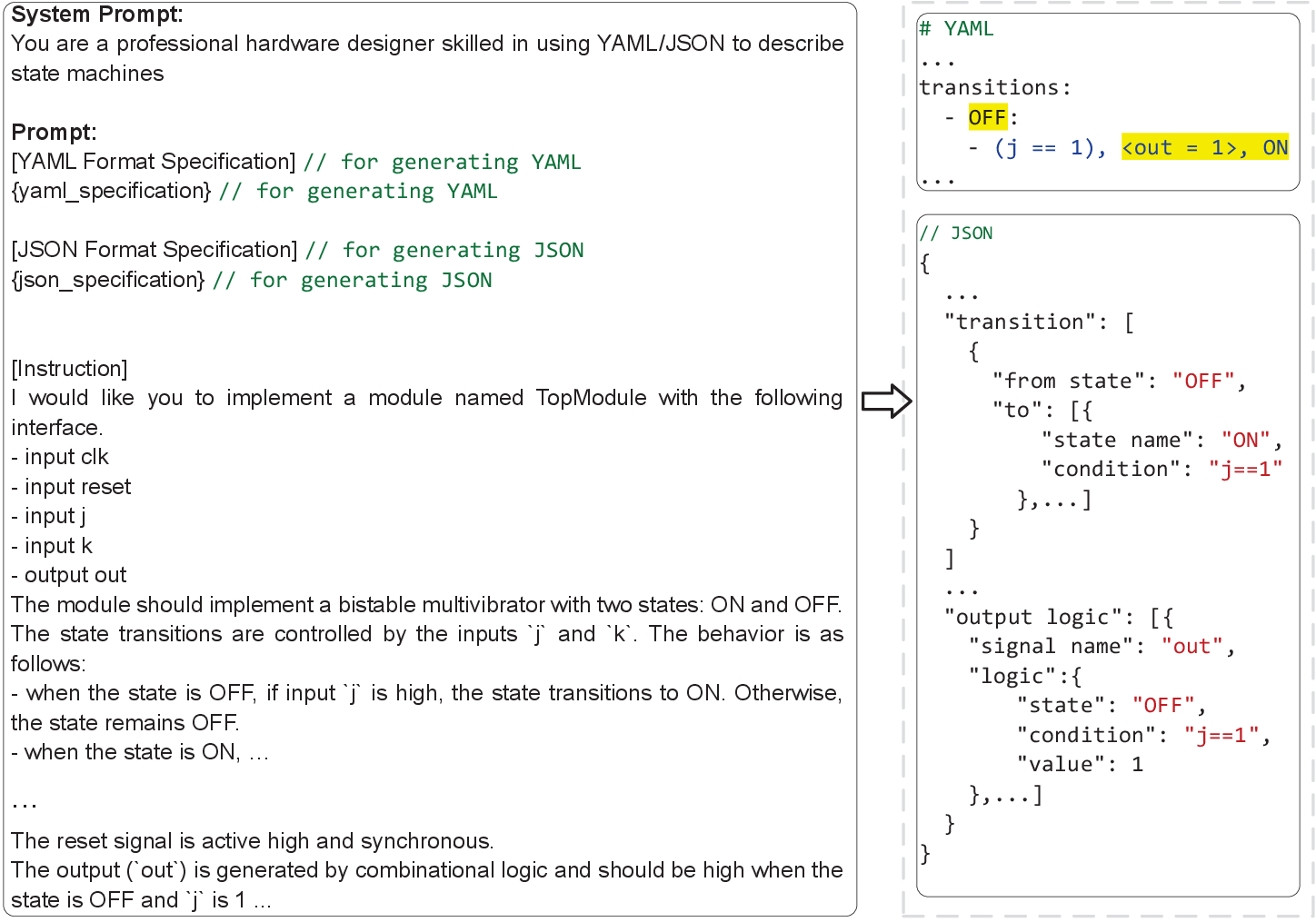}
	\caption{YAML and JSON generation by LLM based on the design description}
	\label{fig:yaml}
\end{figure}

We begin by exploring the LLM's ability to generate YAML-format FSM descriptions compatible with fsm2sv, based on natural language inputs, as illustrated in Fig.~\ref{fig:yaml}. While fsm2sv supports transforming FSMs defined in its specific YAML schema into Verilog code, our experiments revealed several limitations. Due to the compact nature of the YAML schema used by fsm2sv, the LLM sometimes struggles to infer and correctly populate required fields. For example, as shown in the second highlighted region in Fig.~\ref{fig:yaml}, the model mistakenly reversed the positions of “<out = 1>” and “ON”, leading to a misinterpretation by the tool. Additionally, YAML syntax is prone to semantic ambiguity, where keywords or special characters may conflict with signal names or transition conditions, e.g., the name “OFF” is reserved in YAML and results in a syntax error.

To address these issues, we propose using JSON as the intermediate representation format. The LLM is tasked with extracting FSM information from natural language and generating structured JSON output. As shown in Fig.~\ref{fig:yaml}, thanks to its expanded field definitions and stricter syntax, JSON allows the model to more accurately capture and represent the semantic information contained in the input description. This approach significantly reduces errors caused by YAML’s structural ambiguity.

We then develop a json2yaml converter, which transforms the generated JSON into a YAML format that conforms to fsm2sv requirements. This YAML is subsequently passed to fsm2sv to generate the final Verilog code. In addition, we introduce several enhancements to the fsm2sv tool, such as support for customizable clock and reset signal names, the ability to add internal registers beyond the state register, and improve clarity of error messages, all of which extend the applicability and robustness of the overall framework.

As shown in Fig.~\ref{fig:detail}, to ensure consistency between the generated JSON IR and the FSM described in the input task, we introduce a Verifier agent. If inconsistencies are detected, the Verifier reports the issue back to the FSMExtractor, which regenerates the IR. Only JSON outputs that pass verification proceed to the next stage. During the subsequent Verilog generation process, the toolchain may also return FSM generation errors, often due to extra or missing information in the JSON that leads to invalid FSM structures. In such cases, the error is sent back to the FSMExtractor for regeneration. This iterative process continues until either a valid Verilog FSM is successfully generated or the maximum number of iterations is reached.

\subsection{Automatic Testing Based on SystemC and Feedback Optimization}\label{sec:test}
Constructing a reference model and applying identical input stimuli to both the reference and the design-under-test (DUT) to compare their outputs is a widely adopted method for verifying the functional correctness of the DUT. The reference model is typically written in C++, and SystemC, a C++ library for hardware modeling, is often used in such contexts. Therefore, this study employs SystemC for FSM modeling and verification. The SystemC-based verification process involves the following steps: First, a functionally abstract SystemC reference model is constructed. Second, a verification platform is built, incorporating a test stimulus generation module and instantiating both the DUT and the reference model. Finally, co-simulation is performed to compare their outputs and validate the correctness of the design.

\begin{figure}[H]
	\centering
	\includegraphics[width=0.8\textwidth]{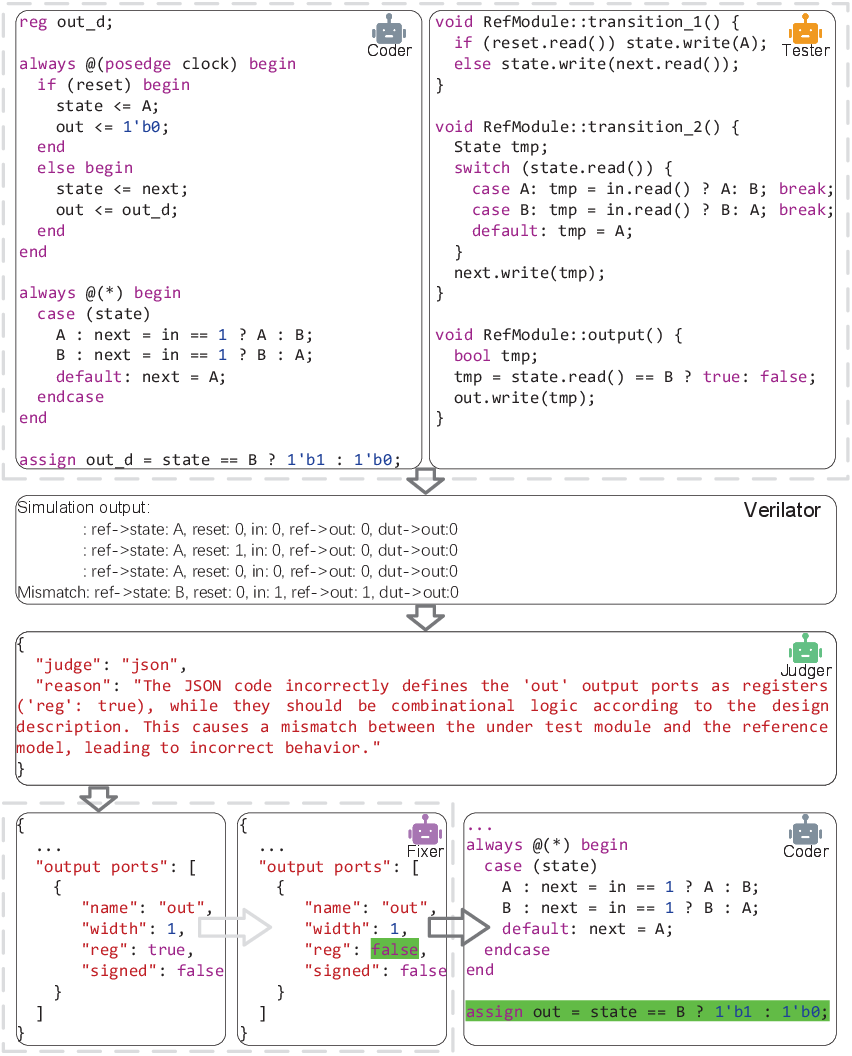}
	\caption{An example of fixing the DUT based on simulation feedback}
	\label{fig:debug}
\end{figure}

To integrate this verification flow into the proposed framework, the generated test program is divided into three parts, including the reference model, the test model (for generating test stimuli), and the main program (which serves as the instantiation and verification entry point). As illustrated in Step 3 of Fig.~\ref{fig:flow}, the verification process begins with the LLM modeling the design description in SystemC and generating the ref.cpp reference model file. The LLM then proceeds to construct the verification platform, producing the test.cpp test model file and the main.cpp main program file. These three files, along with the dut.v file generated in Step 2 (serving as the DUT), enter the simulation and verification phase. The Verilator tool is used to compile the input C++ models. If compilation fails, syntax errors are reported. Only after a successful compilation does the simulation proceed.

The detailed simulation process is illustrated in Fig.~\ref{fig:detail}. The test model randomly generates input stimuli, which are fed simultaneously into both the reference model and the DUT. Their outputs are then compared cycle by cycle. If a mismatch is detected, it indicates a functional error in either the reference model or the DUT, and the simulation is terminated. To assist the LLM in accurately locating the root cause of functional errors, the system records several state-related signals of the reference model prior to the error. This enables the LLM to perform reasoning and localization of the fault. Specifically, in each simulation cycle, the system logs the current state of the reference model, including its inputs, outputs, and register values, as well as the DUT’s outputs, and stores this information in a queue. If the simulation is terminated due to a functional error, all the data in this queue is output as an error trace.

The Judger agent analyzes this error trace to determine whether the fault lies in the reference model or the DUT, and forwards its decision to the Fixer agent. The Fixer then modifies the corresponding model accordingly. Fig.~\ref{fig:debug} shows an example of this process, from simulation to error feedback and ultimately to the correction of the DUT. In this example, the DUT was generated incorrectly, while the reference model was correct. After the simulation failed during Verilator compilation, the Judger agent accurately identified the cause of the error based on the feedback. The Fixer agent then corrected the error in the original JSON file, allowing the Coder agent to regenerate the correct Verilog code.

\section{Experiments}
\subsection{Experimental Setup}
We implement the proposed method based on the MetaGPT \cite{hong2023metagpt} multi-agent framework and conduct experimental evaluation using our self-constructed benchmark, SKT-FSM. As a baseline for comparison, we adopt the current open-source multi-agent code generation framework MAGE, and use outstanding DeepSeek-V3 and GPT-4o as the underlying LLMs for both frameworks. In addition, we conduct additional tests on several well-known LLMs. To ensure the stability and consistency of the results, we set the temperature parameter of the models to 0 and limited the maximum number of agent actions per generation round to 20. Within the AutoFSM framework, the model first generates Verilog code based on the design descriptions in the benchmark. The generated code, along with the corresponding reference model and testbench, is then compiled and executed using Icarus Verilog (iverilog) \cite{williams2002icarus}. Finally, execution results are collected for statistical analysis and evaluation.

\subsection{Evaluation Metrics}

To evaluate the functional correctness of the generated code, we adopt the $\textnormal{pass}@k$ metric, which is commonly used in recent related studies 
\cite{chen2021evaluating,liu2024rtlcoder,ranga2024rtl}. It measures the probability that at least $c$ of $k$ selected candidates pass all test cases, given $n$ candidate code generations per problem. The computation is:
\begin{equation}
\small
\textnormal{pass}@k = \underset{{problems}}{\mathbb{E}}\left[1 - \frac{\binom{n-c}{k}}{\binom{n}{k}}\right]
\end{equation}
In addition, to assess our advantage in ensuring syntactic correctness, we introduce the syntax error rate, defined as follows:
\begin{equation}
\small
syntax\ error\ rate = \underset{{problems}}{\mathbb{E}}\left[\frac{c}{n}\right]
\end{equation}
where $n$ denotes the total number of iterations in Verilog code generation process, and $c$ represents the number of iterations in which syntax errors occurred.

\subsection{Main Results}
Table~\ref{tab:pass-syntax} presents a comparison of AutoFSM, MAGE, and LLMs on the SKT-FSM benchmark regarding pass@1 and syntax error rate. With DeepSeek-V3, AutoFSM outperforms MAGE with an 11.94\% improvement in pass@1 and a 0.49\% reduction in syntax error rate. Under GPT-4o, AutoFSM achieves a 5.97\% higher pass@1 and a significant 17.62\% decrease in syntax error rate. These results indicate that AutoFSM consistently generates higher-quality and more syntactically correct Verilog code compared to MAGE when using the same LLM. Moreover, AutoFSM’s pass@1 performance surpasses that of other LLMs when using both DeepSeek-V3 and GPT-4o. Although GPT-4o’s pass@1 is slightly lower than that of Qwen2.5-Max, its integration with our framework enables it to outperform Qwen2.5-Max, resulting in a 2.99\% improvement in pass@1.

\begin{table}[htbp]
\small
  \centering
  \caption{Pass rates and syntax error rates across LLMs and code generation frameworks}
    \begin{tabular}{lcccc}
    \toprule
    \textbf{Method} & \textbf{Model Size} & \textbf{Model Type} & \textbf{pass@1} & \textbf{Syntax Error Rate} \\
    \midrule
    DeepSeek-V3 & 685B  & Open  & 35.80\% & 14.93\% \\
    MAGE (DeepSeek-V3) & 685B  & Open  & 46.27\% & 1.24\% \\
    AutoFSM (DeepSeek-V3) & 685B  & Open  & \textbf{58.21\%} & \textbf{0.75\%} \\
    \midrule
    GPT-4o & Undisclosed & Closed & 31.34\% & 14.93\% \\
    MAGE (GPT-4o) & Undisclosed & Closed & 38.81\% & 22.84\% \\
    AutoFSM (GPT-4o) & Undisclosed & Closed & \textbf{44.78\%} & \textbf{5.22\%} \\
    \midrule
    Gemini 2.5 & Undisclosed & Closed & 34.33\% & 56.72\% \\
    Qwen2.5-Max & Undisclosed & Closed & 41.79\% & 16.42\% \\
    Qwen-Turbo-0248 & Undisclosed & Closed & 26.87\% & 11.94\% \\
    \bottomrule
    \end{tabular}%
  \label{tab:pass-syntax}%
\end{table}%

\begin{figure}[H]
	\centering
\includegraphics[width=0.9\textwidth]{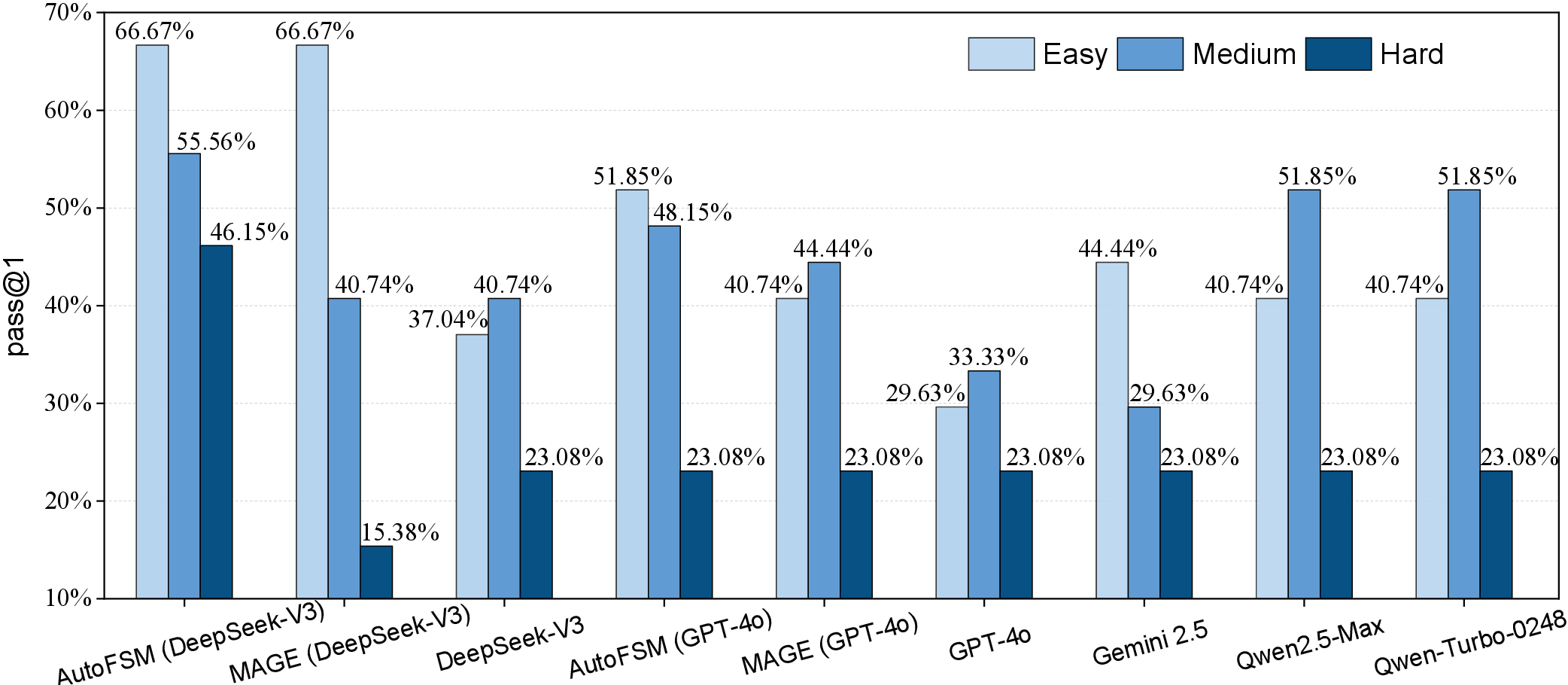}
	\caption{Pass rates of LLMs and code generation frameworks across difficulty levels}
	\label{fig:different-level}
\end{figure}

Fig.~\ref{fig:different-level} further illustrates the pass@1 performance of AutoFSM, MAGE, and LLMs across samples of varying difficulty levels (Easy, Medium, Hard). With DeepSeek-V3, AutoFSM shows a 14.82\% improvement on Medium-level tasks and a notable 30.77\% improvement on Hard tasks over MAGE, demonstrating stronger generalization capabilities for FSM synthesis. Using GPT-4o, AutoFSM also surpasses MAGE across all difficulty levels, improving by 11.11\% on Easy and 3.71\% on Medium samples, further validating its superior generation ability under different model and task conditions.

\subsection{Ablation Study}
To verify the effectiveness of two key design components in AutoFSM, our ablation study focuses on: (1) the use of intermediate representation (IR) and its toolchain to reduce syntax errors, and (2) the automatic generation of test programs and feedback optimization based on SystemC. Given AutoFSM’s superior performance with DeepSeek-V3, the ablation study is conducted on this model.

\begin{table}[htbp]
\small
  \centering
  \caption{Pass rates and syntax error rates after component removal}
    \begin{tabular}{ccc}
    \toprule
    \textbf{Method} & \textbf{{pass@1}} & \textbf{Syntax Error Rate} \\
    \midrule
    AutoFSM without Tester & 46.27\% & 0.00\% \\
    AutoFSM without Tester and JSON & 50.75\% & 11.94\% \\
    DeepSeek-V3 & 35.80\% & 14.93\% \\
    AutoFSM & 58.21\% & 0.75\% \\
    \bottomrule
    \end{tabular}%
  \label{tab:ablation}%
\end{table}%

Table~\ref{tab:ablation} reports the performance impact of removing different components on pass@1 and syntax error rate. When the Tester agent is removed (i.e., test generation and feedback optimization are disabled), pass@1 drops from 58.21\% to 46.27\%, a decline of 11.94\%. Further removing the IR (JSON) and directly using YAML encoding causes pass@1 to drop to 50.75\%, with the syntax error rate rising from 0.75\% to 11.94\%, indicating a substantial degradation. If all key innovations are removed and only the base DeepSeek-V3 model is used, pass@1 drops further to 35.8\%, and the syntax error rate increases to 14.93\%. These results highlight that both the IR and test-feedback mechanism are critical to improving generation quality and reducing syntax errors—serving as core contributors to system performance.

\section{Conclusion and Future Work}
This paper proposes AutoFSM, a multi-agent collaborative framework for the task of FSM generation. By introducing a structured IR to replace the direct generation of Verilog code, AutoFSM significantly reduces syntax error rates (by up to 17.62\%) and enhances both the structural regularity and functional correctness of the generated code. The framework further integrates SystemC-based modeling and automatic test program generation, addressing the limitations of existing methods that rely heavily on predefined testbenches, and providing more interpretable and targeted feedback during debugging. In addition, we develop a new benchmark suite, SKT-FSM, to systematically evaluate FSM code generation performance and provide a unified evaluation standard for future research. Experimental results show that, under the same base LLM, AutoFSM improves the pass rate by up to 11.94\% over the MAGE framework and consistently outperforms the baseline model. Ablation studies further validate the effectiveness of our approach: removing the test generation and IR modules leads to a 22.41\% decrease in pass rate and a 14.18\% increase in syntax error rate, highlighting the critical contributions of these components to the system's overall performance.

In future work, we plan to extend AutoFSM to support a wider range of control logic patterns, such as complex nested state machines. Additionally, we aim to integrate AutoFSM into mainstream EDA toolchains, enabling full-process automation from design specification to hardware deployment.

\begin{credits}
\subsubsection{\ackname}
This work was supported by the Shenzhen Science and Technology Foundation (20250524222348001);  Enterprise R\&D Special Projects of Tianjin Science and Technology Bureau (No. 23YFZXYC00039); Huawei Developer Advocate Program.
\end{credits}

\bibliographystyle{splncs04}
\bibliography{mybibliography}
%




\end{document}